\documentstyle[aps,twocolumn,psfig]{revtex}
\begin{document}
\draft
\title{Tunneling rate fluctuations induced by non-linear resonances: a
quantitative treatment based on semiclassical arguments}
\author{Luca Bonci$^{1}$, Andrea Farusi$^{2}$, Paolo Grigolini$^{1,2,3}$ 
and Roberto
Roncaglia$^{4} $}
\address{$^{1}$Center for Nonlinear Science, University of North Texas,
P.O. Box 5368, Denton, Texas 76203 }
\address{$^{2}$Dipartimento di Fisica dell'Universit\`{a} di Pisa, Piazza
Torricelli 2, 56127 Pisa, Italy }
\address{$^{3}$Istituto di Biofisica del Consiglio Nazionale delle
Ricerche, Via San Lorenzo 26, 56127 Pisa, Italy }
\address{$^{4}$FORTH, Institute of Electronic Structure and Laser,
P.O. Box 1527, 711 10 Heraclion, Cretre, Greece}
\date{\today}
\maketitle

\begin{abstract}
We investigate the tunneling process between two symmetric stable islands 
of a forced pendulum Hamiltonian in the weak chaos regime. We show that 
when the tunneling doublet is quantized over a classical non-linear 
resonance the tunneling rate strongly deviates from the semiclassical 
prediction. This mechanism is responsible for the irregular dependence of 
the tunneling rate on the system parameters. The weak-chaos condition 
allows us to make a theoretical prediction which agrees very well with the 
numerical results. This opens up a possible avenue to a general theory on 
the dependence of quantum tunneling on classical chaos.
\end{abstract}

One of the most attractive aspects of classically chaotic quantum systems 
is the surprising behavior of the tunneling rate. This is shown to be an 
erratic function of the system parameters~\cite{lb,btu93,prlnoi}. This 
means that the tunneling rate undergoes fluctuations which enhance, or 
reduce, its intensity by several orders of magnitude, compared to the 
smooth and regular behavior corresponding to the traditional conditions.

A kind of general agreement has been reached by the researchers working on 
this hot issue. This is that the tunneling properties must be traced back 
to the crossing of the tunneling doublet with a third state~\cite{D97}, 
which corresponds to the chaotic region of the classical phase 
space~\cite{btu93}. We shall be referring to these phenomena as Chaos 
Assisted Tunneling (CAT) processes, since this is the definition generally 
adopted in literature in spite of the fact that the influence of chaos can 
also reduce the rate of the tunnel process. It is also thought, at the 
level of merely qualitative arguments, that the erratic-like behavior of 
the tunneling rate depends on the chaotic nature of the third state. It 
has to be pointed out that a semiclassical treatment represents the natural 
way of establishing a connection between the quantum properties, tunneling 
rate in the case under discussion, and the classical phase space.  
Unfortunately, these methods cannot be applied to the chaotic states, and 
this is probably the reason why all the results reached so far on this 
interesting issue fail supplementing the qualitative arguments with precise 
quantitative predictions.

This letter is devoted to describing the discovery that the processes 
observed in literature so far, the CAT processes, are a special case of a 
more general phenomenon. The third state invoked by all the researchers of 
this field need not be chaotic. Here we show that the same qualitative 
behavior as that observed in \cite{lb,btu93} becomes ostensible also as an 
effect of the crossing with regular states. In this case the crossings of 
states, and thus the fluctuation of the tunneling doublets, can be directly 
related to the classical phase-space structure and in particular to the 
destruction of the regular tori by the non-linear resonances. This makes 
it possible for us to derive a quantitative prediction, which turns out to 
agree remarkably well with the numerical results.

\begin{figure}[tbp]
\centerline{\psfig{figure=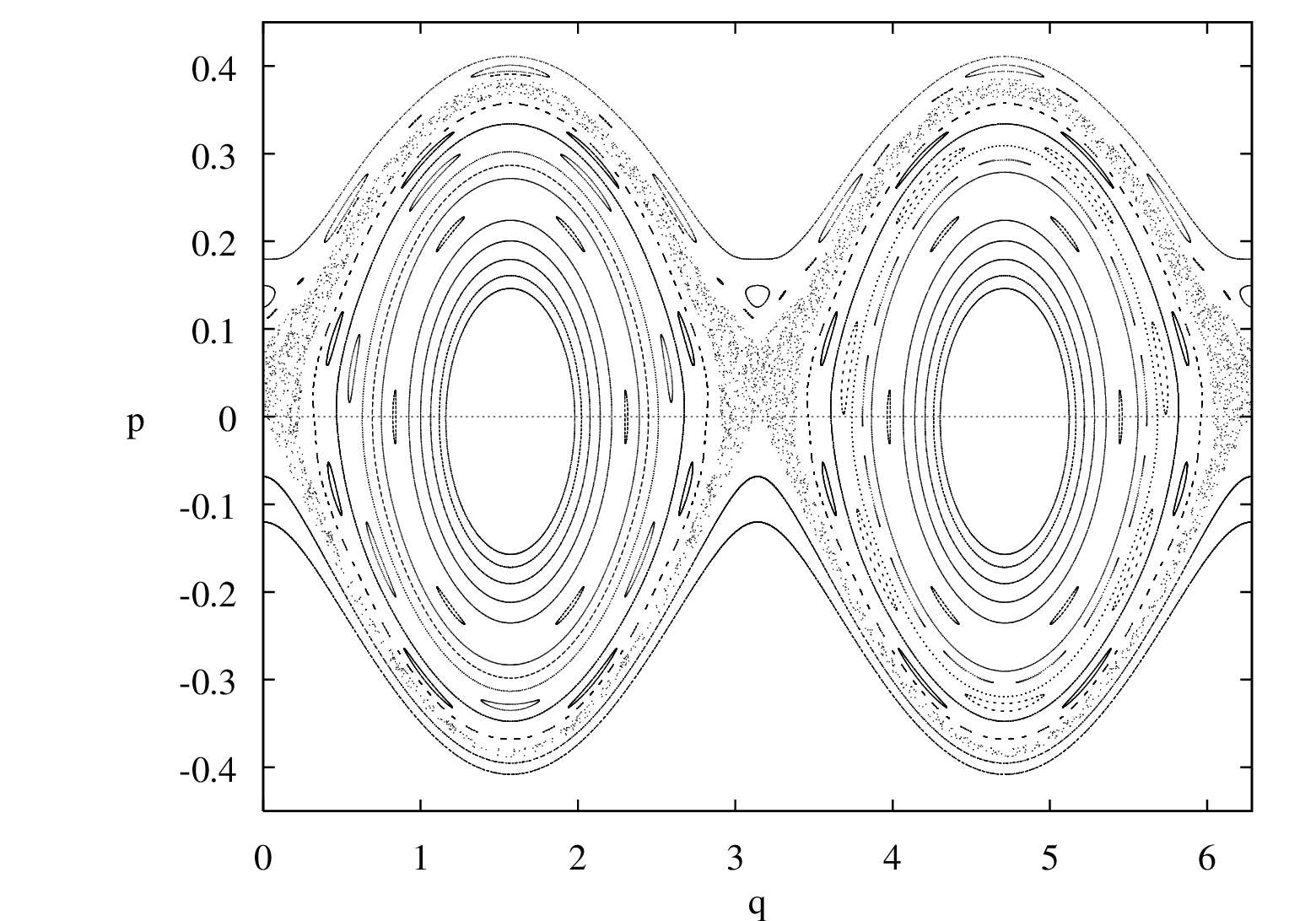,width=2.6in}}
\caption[]{Stroboscopic plot of the classical dynamics driven by 
Hamiltonian
(\ref{hamiltonian}). The values of the system parameters are $\Omega =2$, 
$v_1=0.035$, $v_2=0.005$. Some non-linear resonances of order $1/n$ are 
visible: those below the separatrix refer to $n$=6, 7 and 8, and that 
above it to $n$=8.}
\label{fig:class}
\end{figure}

To address the problem of the connection between tunneling and classical
phase-space structure we consider the following Hamiltonian 
\begin{equation}
H=\frac{p^{2}}{2}+v_{1}\cos (2q)+v_{2}\cos (2q-\Omega t)\ .
\label{hamiltonian}
\end{equation}
This is a forced system whose dynamical properties are well known.  
Depending on the value of the perturbation strength $v_{2}$, the phase 
space can show either regular or chaotic dynamics. For $v_{2}=0$ this is 
nothing but an ordinary, and integrable, pendulum. Increasing the value of 
$v_{2}$ makes the dynamics non integrable and eventually chaotic.  
Fig.~\ref{fig:class} shows the effect of setting $v_{2}=0.005.$ We can 
easily identify the isolated resonances of order $1/8,\ 1/7$ and $1/6$, 
resulting from the destruction of tori with winding numbers rationally 
connected to the perturbation frequency. Notice also the stochastic layer 
close to the separatrix, where the higher-order resonances overlap, giving 
rise to fully developed chaos.

To study the quantum dynamics of this system, and in particular its 
tunneling properties, we set periodic conditions at the borders of the 
interval $q\in [0,2\pi ]$. This has the effect of creating an enlarged 
phase space consisting of two identical cells. The quantum mechanical 
eigenstates must be either symmetric or anti-symmetric with respect to a 
$\pi$-translation along the $q$-axis, and some eigenstates exist whose 
linear combination is essentially located in only one of the two cells.  
According to~\cite{floq} the time evolution is described by means of a 
unitary operator, which evolves the quantum state by one entire period of 
the external perturbation. This is called Floquet operator and it is 
denoted by $\hat{F}$. The generalized eigenstates $u_{n}(q,t)$ and 
eigenvalues $\varepsilon_{n}$ of $\hat{F}$ are derived from the eigenvalue 
equation:
\begin{equation}
(H-i\hbar \frac{\partial }{\partial t})u_{n}(q,t)=\varepsilon
_{n}u_{n}(q,t)\ ,  \label{eigenfloq}
\end{equation}
where $u_{n}(q,t)=\exp (-i\phi _{n}t)\psi _{\phi _{n}}(q,t)$ is by 
construction time periodic, $u_{n}(q,t)=u_{n}(q,t+T)$, and $\varepsilon 
_{n}$ is defined by $\varepsilon _{n}\equiv\hbar \phi _{n}$.  One important 
feature of the Floquet formalism is the Brillouin-zone structure of the 
spectrum: every eigenstate $u_{n}(q,t)$ results in the wider class of 
eigenstates and eigenvalues defined respectively by
\begin{equation}
u_{n,m}(q,t)\equiv u_{n}(q,t)e^{im\Omega t}  \label{stabrillo}
\end{equation}
and
\begin{equation}
\varepsilon _{n,m}\equiv \varepsilon _{n}+\hbar m\Omega \ ,  
\label{enbrillo}
\end{equation}
where $m\in Z$. It is evident that the functions belonging to the same 
class represent the same solution of Eq.~(\ref{eigenfloq}). Therefore, if 
a eigenstate has a given eigenvalue $\varepsilon $, which does not belong 
to the first Brillouin region, $0\leq \varepsilon <\hbar \Omega $, it is 
convenient to fold it back to this region by means of the prescription 
(\ref{enbrillo}). To avoid confusion we denote the resulting energies 
with the symbol $E$.

The eigenstates $u_{n,m}(q,t)$ reflect the translational invariance $ 
q\rightarrow q+\pi $ and can be labelled as odd or even with respect to 
this symmetry operation. Thus the energy spectrum of the Floquet operator 
appears as a series of doublets $E_{n,\pm }$ with $\pm $ indicating the 
translation simmetry. The rate of the tunneling process is determined by 
the energy splittings $\Delta E_{n}=\left| E_{n,+}-E_{n,-}\right| $ which, 
in turn, determine the dynamics of a generic wavepacket initially located 
in one of the two islands. For simplicity here we focus on the energy 
splitting of the ``main doublet'', which represents the natural extension 
of the fundamental-state doublet of an autonomous Hamiltonian to the 
Floquet picture. This is defined as the set of the two eigenstates of the 
Floquet operator that have the largest overlap with a minimum-uncertainty 
state located at the center of one of the two stable islands~\cite{prlnoi}.

\begin{figure}[tbp]
\centerline{\psfig{figure=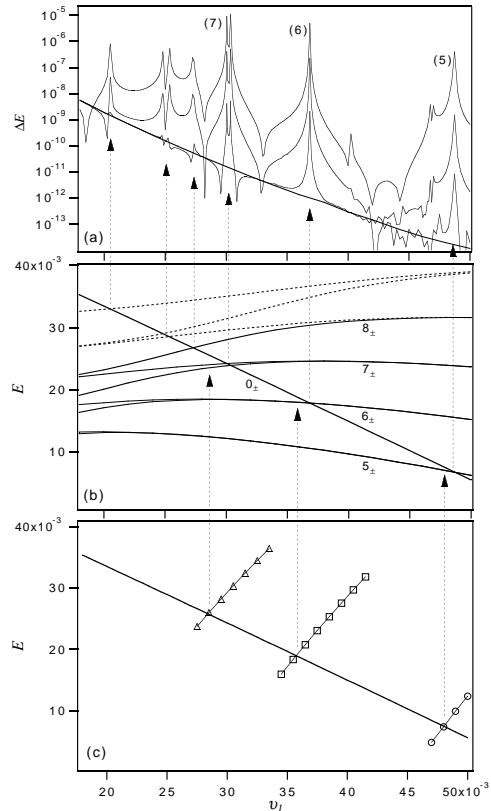,width=2.6in}}
\caption[]{The origin of the tunneling rate fluctuations for 
Hamiltonian~(\ref{hamiltonian}) with $\Omega=2$ and $\hbar= 0.025$.
(a) The main doublet splitting $\Delta E$ as a function of $v_1$. The thick
full line denotes the unperturbed case $(v_2 = 0)$. The other curves, from 
the bottom to the top, refer to $v_2 = 0.0001$, $0.001$ and $0.005$.
(b) The levels $E_n$ as functions of $v_1$.
The full line labelled by $0\pm$ denotes the main doublet.
The other levels are indicated by either full or dashed lines according to 
whether their crossing with the main doublet occurs below or above the 
separatrix. The full lines are labelled by their quantum numbers. The value
of the perturbation strength is $v_2=0.001$. All the energy splittings, also
at the avoided crossings, are not visible on this scale.
(c) The energy of the fundamental doublet (full line) and the energies
of the renormalized quantum resonances as solutions of 
Eq.~(\ref{resocor2}) (symbols). The resonances are of order $1/n = 1/5$ 
(circles), $1/6$ (squares) and $1/7$ (triangles).
The vertical arrows are eye-guides connecting the theoretical
crossings of (c) to the numerical crossings of (b), and these to the
tunneling peaks of (a).}
\label{fig:fig2PRL}
\end{figure}

We adopt the following numerical procedure. The Floquet matrix $\hat{F}$ 
is determined by the numerical integration of~(\ref{eigenfloq}) and the 
eigenvalues and the eigenstates are subsequently obtained by numerical 
diagonalization. The results for the main-doublet splitting $ \Delta 
E\equiv \Delta E_0$, as a function of $v_{1}$ are shown in 
Fig.~\ref{fig:fig2PRL}a for different values of the perturbation parameter 
$v_{2}$.  We see that accordingly with the semiclassical 
prediction~\cite{w86}, in the unperturbed case $\Delta E$ is a smooth 
function of $v_{1}$. However, when $v_{2}$ is given an even small but 
finite value, $\Delta E$ strongly departs from the smooth behavior, with an 
increase, or decrease, by several orders of magnitude. This behavior is 
similar to that of~\cite{btu93}, where the splitting irregularity has been 
related to the crossings of the tunneling doublet with a third level which 
belongs to the chaotic region of the corresponding classical system.  To 
confirm the connection of these results with the third-state theory, we 
calculate the first energy levels $E_n$. In Fig.~\ref{fig:fig2PRL}b we 
show those of them which cross the main doublet in the same 
$v_{1}-$interval as that of Fig.~\ref{fig:fig2PRL}a. We see clearly that 
the splitting irregularities correspond to the crossing between the main 
doublet and a third state, or more precisely, a second doublet~\cite {D97}.  
The effect of level crossing with a weakly perturbed Hamiltonian is well 
established: the spectrum is modified only in the vicinity of the crossing 
which becomes an avoided level crossing, thereby resulting in significant 
changes of the main-doublet levels, and thus of the splitting $ \Delta E$.  
Note that the avoided crossing is too small to be visible in the scale of 
Fig~\ref{fig:fig2PRL}b.

Is this effect a CAT process? Do the states colliding with the main doublet 
belong to the chaotic region of the phase space? The answer is 
incontrovertible: in general they do not. This is so because, at least for 
the smallest values of $v_{2}$ used in Fig.~\ref{fig:fig2PRL}a, the 
doublets 5, 6 and 7 belong to the regular region. In fact $v_{2}=0.0001$, 
a value at which the fluctuations of the tunneling rate are already very 
strong, means a perturbation weaker than that used to derive 
Fig.~\ref{fig:class}: the corresponding phase space is even more regular, 
and thus barely distinguishable from the unperturbed one except possibly 
for a thin stochastic layer around the separatrix. We think, therefore, 
that the results of Fig~\ref{fig:fig2PRL}a, with the help of 
Fig.~\ref{fig:class}, afford a compelling numerical evidence that the 
tunneling rate can be characterized by strong fluctuations without 
necessarily involving the interaction with a strongly chaotic region. The 
splitting irregularities are shown below to be caused by the birth, in the 
regular phase-space region, of non-linear resonances, a fact that, 
surprisingly enough, has been overlooked by the literature on this field of 
research.

Note that further progress on the CAT processes has been hampered by a 
major difficulty: it is not yet known how to realize a proper semiclassical 
picture of the chaotic states and, consequently, how to evaluate the 
tunneling matrix elements. The crossings 5, 6 and 7 of 
Fig.~\ref{fig:fig2PRL}a, on the contrary, are compatible with the 
semiclassical quantization of Hamiltonian (\ref{hamiltonian}), along the 
lines established by Breuer and Holthaus~\cite{breuer}. The analysis here 
is restricted to the crossing between the main doublet and the states lying 
below the separatrix.

We proceed as follows. Adopting a perturbative approach, we replace $H$ 
with $H_{0}=\frac{p^{2}}{2}+v_{1}\cos (2q)$, and write the condition of 
level crossing in the Brillouin zone as~\cite{breuer}:
\begin{equation}
H_{0}(\hbar (n+\frac{1}{2}))+\hbar \Omega m=H_{0}(\hbar (n^{\prime 
}+\frac{1}{2}))+\hbar \Omega m^{\prime }.  \label{degenera}
\end{equation}
Let us assume $\hbar $ to be so small as to make it possible to consider 
$\hbar (n-n^{\prime })$ to be a small expansion parameter. By using $J=\hbar
(n+1/2)$, we obtain 
\begin{equation}
\overline{\omega }=\frac{\Delta m}{\Delta n}\Omega +{\cal O}(\hbar )\ ,
\label{resonance}
\end{equation}
where $\overline{\omega }\equiv \overline{\omega }(J)=\frac{\partial 
H_{0}}{\partial J}$ is the frequency of the unperturbed libration as a 
function of the classical action $J$, $\Delta n$ $\equiv n^{\prime }-n$ and 
$\Delta m\equiv $ $m-m^{\prime }$. Eq.~(\ref{resonance}) corresponds to 
the classical condition for the onset of non-linear resonances. This means 
that, for sufficiently small $\hbar $'s, the crossing of the two 
unperturbed levels $E_{n,m}^{0}$ and $E_{n^{\prime },m^{\prime }}^{0}$ 
occurs as a quantum reflection of the birth of a non-linear resonance of 
order $\Delta m/\Delta n$ in the classical phase space. The latter has the 
same energy as the torus determining the semiclassical quantization of one 
of the two crossing levels.  On the basis of this result the level crossing 
concerning the main doublet $(n=0)$ corresponds to the overlap between the 
semiclassical quantization torus of the fundamental state and a non-linear 
resonance of the appropriate order. Notice that, in principle, Eq.~(\ref 
{resonance}) involves resonances of any order thereby making a given level 
undergo a virtually infinite number of crossings upon change of a system 
parameter. However, Ref.~\cite{breuer} shows that the perturbation 
strength at the crossing is determined by the order of the corresponding 
non-linear resonance, and that the first-order resonances ($\Delta m=1$) 
produce the most intense effects.

We have now to point out that our numerical results are confined to finite, 
and relatively large, values of $\hbar $ by the obvious limitations of 
computer calculations. This does not conflict with physical reality, since 
an experimentally observable tunnel process implies the Planck constant to 
be finite rather than infinitelly small. In conclusion, to derive a more 
realistic prediction we must go beyond Eq.~(\ref{resonance}), and this, at 
the same time, makes it possible, as we shall see, a reliable comparison 
between the semiclassical prediction and the numerical results. This 
purpose is realized by applying to Eq.~(\ref{degenera}) the second-order 
approximation. We obtain:
\begin{equation}
\overline{\omega }=\frac{\Delta m}{\Delta n}\frac{\Omega }{(1+\frac{\hbar 
}{2}\frac{d\overline{\omega }}{dE}\Delta n)}+{\cal O}(\hbar ^{2}),
\label{resocor2}
\end{equation}
where $\frac{d\overline{\omega }}{dE}$ can be expressed analitically via 
$\overline{\omega }=\pi \sqrt{v_{1}}/K(k)$, $K(k)$ is the elliptic function
and $k^{2}=(E+v_{1})/2v_{1}$. Eq.~(\ref{resocor2}) looks like Eq.~(\ref
{resonance}) with the frequency $\Omega $ properly renormalized and thus
depending on $\hbar $, as well as on the energy of the colliding level.

This theoretical prediction on the quantum crossings is illustrated by 
Fig.~\ref{fig:fig2PRL}c. We note that the fundamental doublet energy is 
denoted by a single full line because, as in Fig.~\ref{fig:fig2PRL}b,
the energy splitting is not visible in this scale.
The prediction stemming from 
Eq.~(\ref{resocor2}) must be compared to the numerical quantum crossings of 
Fig.~\ref{fig:fig2PRL}b, which occur both above and below the separatrix.  
Here we see that the accuracy of the prediction is good and, as expected, 
becomes better with resonances of smaller order. A still better agreement 
is expected from smaller values of $\hbar $.

We now invite the reader to compare \ref{fig:fig2PRL}b to 
\ref{fig:fig2PRL}a, while keeping in mind its earlier correspondence with 
\ref{fig:fig2PRL}c. This results in a vivid explanation of the 
``fluctuations'' of the tunneling rate. This is made especially evident by 
a thorough examination moving from smaller to larger values of the 
perturbation strength $v_{2}$.  We see that at $v_{2}=0.0001$ the 
deviations from the unperturbed behavior are 
significant only in the close vicinity of the level crossings, thus lending 
support to our perturbative approach. With the increase of $v_{2}$, the 
peaks broaden and overlap one another and for still larger values of 
$v_{2}$ new peaks come forth corresponding to crossings of higher order.

The peaks around $v_{1}=0.025$ and $v_{1}=0.0275$ refer to levels close to 
the separatrix, which at properly large $v$'s are embedded in the chaotic 
region. The peak at about $v_{1}=0.020$ refers to an above-separatrix 
level. As a consequence of this, the dynamical classical process behind 
these three peaks is compatible with the motion from the one to the other 
potential well. For these reasons one would be tempted to conclude that 
the corresponding transition rates are the largest. We see, on the 
contrary, that the peaks referring to a below-separatrix level crossing 
($v_{2}>0.030$) are more intense than the peaks due to the above-separatrix 
level crossings. We also note that at the highest value of $v_{2}$ used in 
Fig.~\ref{fig:fig2PRL}a, when the tunneling rate results in the maximum 
departure from the unperturbed prediction, the classical phase-space 
structure is still almost regular, as shown by Fig.~\ref{fig:class}, which 
refers to the same value of $v_{2}$.

In conclusion, we have identified a process resting on the role of
isolated non-linear resonances rather than that of the connected chaotic
sea. The corresponding tunneling rate $\Delta E$ exhibits fluctuations which
can be even more intense than those provoked by the crossing with chaotic
states. This can be accounted for by using the same heuristic arguments as
those adopted in Ref.~\cite{btu93}. These authors argue that the intensity
of the process is proportional to the matrix element $V$
connecting one of the partner of the tunneling doublet, state
$|r\rangle$, to the crossing state $|c\rangle$. This crucial parameter is
roughly proportional to the overlap between $|c\rangle$ and $|r\rangle$.
We can predict its value with heuristic arguments based on the observation
of the Husimi distribution of the eigenstates (not shown here).
The ground state is represented by almost gaussian packets located at
the centers of the regular islands, the regular crossing state corresponds
to a double ring-shaped bun surrounding the gaussian packets,
and the chaotic state to an eight-shaped distribution lying on the chaotic
region of Fig.~\ref{fig:class}.
As the value of $\hbar$ decreases, the gaussian state shrinks,
the regular distribution collapses on its quantization torus and the
chaotic state remains localized in the chaotic layer. This produces a
decrease of the overlap among different states, and consequently an overall
decrease of the tunneling rate. However, a mere inspection of 
Fig.~\ref{fig:class} leads to the incontrovertible conclusion that
the effect of the regular-regular crossing is expected to remain larger
than that of the regular-chaotic one. This is made compelling by the
topology of the phase space: the overlap with the closer regular 
state remains larger than the overlap with the farther chaotic one.

We note also that the values of the parameter $v_{1}$ corresponding
to the tunneling-rate peaks, appear to be independent of the intensity
of the perturbation strength $v_{2}$.
Our conjecture is that the position of the peaks remains 
unchanged even when the overlap of the isolated resonances occurs, and a 
condition of full chaos develops. This changed condition produces the 
broadening of these peaks, the birth of higher-order crossings and the 
resulting merging of all of them into a single erratic-like structure, 
without influencing their original position. Thus we think that 
even after the crossover to the chaotic regime, significant signs of the 
peaks originated by the quasi-integrable non-linear resonances remain.  

Before concluding, we want to point out that these results are
qualitatively similar to those obtained in Ref.~\cite{prlnoi}. These
physical effects, as here shown, can be accounted for very well, both
qualitatively and quantitatively. This means that the theoretical arguments
of this letter provides an exhaustive account for the
``plateau effect'' of~\cite{prlnoi}. This is an important task
which has recently challenged the efforts of some groups (see
Ref.~\cite{zgl95}). Should the phenomenon illustrated
in~\cite{prlnoi} be proved to be a form of CAT, this would be
a strong support of our conviction that the semiclassical arguments of this
letter are a significant step ahead towards a quantitative prediction
of CAT processes.

{\bf Acknowledgments} R.R. thanks the CEE for a postdoctoral fellowship
(TMR) Programme Research Training Grant Nr ERB4001GT952681, while L.B.
acknowledges the Italian CNR and NATO for financial support.


\begin{references}
\bibitem{D97}  The expression ``crossing with a third state'' is borrowed
from the literature, but it is probably inappropriate, since
the third state can also consists of a doublet with a so small
energy separation as to prevent the possibility
of detecting a crossing at a time.
\end{references}
\end{document}